\documentclass[twocolumn,english,prx,superscriptaddress,longbibliography]{revtex4-1}
\usepackage{mathrsfs}
\usepackage{epsfig}
\usepackage{graphicx}
\usepackage{amsfonts}
\usepackage[figuresright]{rotating}
\usepackage{amssymb}
\usepackage{amsmath}
\usepackage{dcolumn}
\usepackage{bm}
\usepackage{color}

\def\Re{\rm{Re}}

\def\be{\begin{equation}} \def\ee{\end{equation}}
\def\bea{\begin{eqnarray}} \def\eea{\end{eqnarray}}

\newcommand{\WQCASQC} {Wilczek Quantum Center and Key Laboratory of Artificial Structures and Quantum Control, School of Physics and Astronomy, Shanghai Jiao Tong University, Shanghai 200240, China}

\newcommand{\wqctdli} {Wilczek Quantum Center, School of Physics and Astronomy and T. D. Lee Institute,
Shanghai Jiao Tong University, Shanghai 200240, China}

\newcommand{\Pittsburgh} {Department of Physics and Astronomy, University of Pittsburgh, Pittsburgh, Pennsylvania 15260, USA}

\begin{document}
\title{Imaginary time crystal of thermal quantum matter}

\author{Zi Cai}
\email{zcai@sjtu.edu.cn}

\author{Yizhen Huang}

\affiliation{\WQCASQC}

\author{W. Vincent Liu}
\email{w.v.liu@icloud.com}
\affiliation{\Pittsburgh}
\affiliation{\wqctdli}

\begin{abstract}


Temperature is a fundamental thermodynamic variable for matter. Physical observables are often found to either increase or decrease with it, or show a non-monotonic dependence with peaks signaling underlying phase transitions or anomalies. Statistical field theory has established connection between temperature and time:  a quantum ensemble  with inverse temperature $\beta$ is formally equivalent to a dynamic system evolving along an imaginary time from 0 to $i\beta$ in the space one dimension higher.  Here we report that a gas of hard-core bosons interacting with a thermal bath manifests an unexpected temperature-periodic  oscillation  of its macroscopic  observables, arising from the microscopic origin of space-time locked translational symmetry breaking and crystalline ordering. Such a temperature crystal, supported by Quantum Monte Carlo simulation,  generalizes the concept of purely spatial density-wave order by mapping the  time axis for Euclidean action to an extra space dimension for free energy.
  \end{abstract}

\maketitle
\section{\bf Introduction}
The phases of quantum matter are, in general, categorized according to symmetries and their spontaneous breaking patterns.
Recently, a striking possibility was put forward~\cite{Wilczek2012},  dubbed ``time crystal'', where the interacting particles spontaneously organize themselves into a ``time'' coherent  state that breaks temporal translational symmetry. Temporal periodicities spontaneously emerge analogous to spatially ordered crystals. This phase has raised considerable interest~\cite{Shapere2012,Li2012,Wilczek2013,Sacha2015,Else2016,Khemani2016,Yao2017,Syrwid2017,Khemani2017,Russomanno2017,Gong2018,Huang2018,Sacha2018,Iemini2018,Chew2019}, and was subsequently proven forbidden in  thermodynamic equilibrium states~\cite{Bruno2013,Watanabe2015}. A variant of the phase was proposed  ~\cite{Else2016,Khemani2016,Yao2017} by considering certain non-equilibrium condition, and has quickly been observed in experiments~\cite{Choi2017,Zhang2017}.

Statistical field theory shows that a $d$-dimensional quantum thermal ensemble is equivalently described by the path integral representing the partition function in $d+1$ Euclidean space where time is imaginary. In the field theory dual description, the thermal distribution weighted by the well-known temperature dependent factor is fully captured by integration over auxiliary imaginary time (``iTime'')  ranging from 0 to $\beta=1/k_B T$ ($k_B$ the Boltzmann constant). In this manner, finite-temperature physics is mapped to the dynamics of Euclidean action.  Motivated by such a profound relation between temperature $T$ and iTime, Wilczek had speculated the novel possibility  of spontaneous translational symmetry breaking in iTime in the end of his original paper, dubbed as ``imaginary time crystal'' (iTC)~\cite{Wilczek2012}.  One may wonder what class of thermal ensembles can show crystalline order in iTime space, whether the ``time'' analogue of phonons exist and what novel effects should manifest in ground state and thermal response functions.


Here, we find a class of open quantum ensembles that shows the iTC phase. A key feature of our mechanism is a non-monotonic bath-induced retarded interaction, whose dependence is on the relative time in two body scattering. That satisfies the two conditions what previously might have seem conflicting, i.e., being time dependent but fully translational-invariant. As we will show in the following, integrating out certain engineered local thermal baths yields such a retarded interaction,  which favors crystalline patterns in iTime with periods determined by the position of interacting potential minimum.  Thermodynamic quantities of the iTC exhibit a striking  behavior of oscillation in temperature, which provides an experimental observable effect for this novel phase.

\begin{figure}[htb]
\includegraphics[width=0.99\linewidth,bb=1 1 770 394]{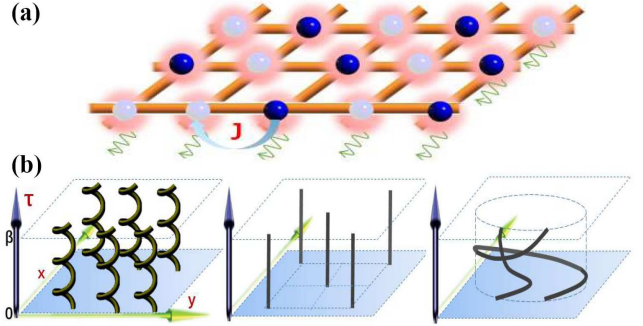}
\caption{(a) A hard-core bosonic lattice model uniformly
coupled to onsite baths.  (b)  Three typical world-line configurations (from left to right) of iTC, Mott-insulator and superfluid phases in a 2+1D Euclidean space.}
\label{fig:fig1}
\end{figure}

\section{\bf Model and method.}
We study  hard-core bosonic models in one-dimensional (1D) lattice of $L$ sites as well as a $L\times L$ square lattice with system Hamiltonian
$H_s=\sum_{\langle ij\rangle} -J[a_i^\dag a_j+a_j^\dagger a_i]$
where $J$ denotes the hopping amplitude and $\langle ij\rangle$ indicates a pair of adjacent sites.  We focus on the half-filling case throughout this paper. On each site $i$ the density operator of the hard-core boson $n_i$ additionally couples to  a local bath, as shown in  Fig.~\ref{fig:fig1}(a).   We assume the total system (system+bath) is in thermodynamic equilibrium with  temperature $T$.  The partition function takes the form $Z_{tot}={\rm Tr}_{s}{\rm Tr}_{b} e^{-\beta H_{tot}}$ (${\rm Tr}_{s/b}$ denotes tracing over the system/bath degree of freedom).  We first integrate out the bath degrees of freedom, which yields a retarded density-density interaction term in iTime.  Thus the open system can be described by the reduced density matrix
\begin{small}
\begin{equation}
\hat{\rho}_{s}= \frac{1}{Z_s}e^{-\beta H_s-S_R}, \label{eq:partition}
\end{equation}
\end{small}
where $H_s$ is the system Hamiltonian and $Z_s={\rm Tr}_{s}\hat{\rho}_s$.   $S_R$ describes
effective action of the translational-invariant onsite bath-induced retarded interaction:
\begin{small}
\begin{equation}
S_R =-\int_0^\beta d\tau \int_0^\beta d\tau' \sum_i
(n_i (\tau) - n_0) D(\tau-\tau') (n_i(\tau') - n_0). \label{eq:ret}
\end{equation}
\end{small}
where $n_0=\frac 12$ is the average particle density, and the site-independent kernel function $D(\tau-\tau')$ depends on the bath properties. For instance,  a  harmonic oscillator bath with frequency $\omega_0$ gives rise to the kernel function  $D(\tau-\tau')\sim e^{-\omega_0|\tau-\tau'|}+e^{-\omega_0(\beta-|\tau-\tau'|)}$ for large $\beta$.  We will show later that the iTC phase can exist for a broad class of kernel functions; here, we chose:
\begin{eqnarray}
D(\tau-\tau') &=&  \alpha [F(\tau-\tau')+F(\beta-|\tau-\tau'|)]  \nonumber \\
F(\tau-\tau') & =& e^{-\omega_d|\tau-\tau'|}\cos2\pi\omega_c (\tau-\tau') \label{eq:interaction}
\end{eqnarray}
where $|\tau|\leq\beta$, and $\alpha$ is the retarded interaction strength.


 The model with the retarded interactions defined in Eq.~(\ref{eq:interaction}) can be more than a toy model of merely academic interest. In general, the interaction induced by baths composed of free bosons in thermal equilibrium is usually attractive, while a fermionic bath-induced interaction is more complex as sometimes it can be oscillating between attraction and repulsion (e.g.the Ruderman-Kittel-Kasuya-Yosida (RKKY) interaction~\cite{Ruderman1954,Kasuya1956,Yosida1957}). Now we need a RKKY-like interaction, but with oscillating decay in iTime instead of real space.  Consider $D(\tau)=\sum_m \frac {e^{i\omega_m\tau}}{i\omega_m-\Sigma_b(i\omega_m)}$ with $\omega_m=\frac{2\pi m}{\beta}$ the Matsubara frequency and $\Sigma_b(i\omega_m)$ the self-energy of the bath.  In general, an imaginary part in self-energy leads to an exponential decay of the quasi-particle in real time (finite lifetime), which corresponds to an oscillation in the imaginary time after analytic continuation. Motivated by Kozii and Fu's recent work about an effective non-Hermitian description for fermionic quasiparticles~\cite{Kozii2017},  In Appendix~\ref{sec:bath}, we propose  a similar microscopic bath model that can induce the oscillating retarded interaction as shown in Eq.~(\ref{eq:interaction}). Moreover, by analogy with interaction in conventional crystals ({\it e.g.} Van der Waals force),  we should emphasize that the time-oscillating feature in the interaction is not a necessary condition for the iTC phase. In Appendix~\ref{sec:alternative}, we have chosen a different form of retarded interaction and numerically shown that the phase can actually exist as long as the retarded interaction is nonlocal and non-monotonic in itime with at least one minimum, whose position defines the ``time'' lattice constant of iTC.

 Path integral Quantum Monte Carlo (QMC) simulation is a stochastic numerical method to study the equilibrium properties of quantum many-body systems based on importance sampling of the world line configurations in Euclidean spacetime.  Following the algorithm proposed in Ref.~\cite{Cai2014}, we generalize the QMC simulation with worm-type updates~\cite{Prokofev1998,Pollet2005} to study the open systems whose distribution functions [Eq.~(\ref{eq:partition})] deviate from the Boltzmann distribution. The key point is to evaluate the integrals resulting from the retardation and include them into the QMC acceptance ratio during the updates  of sampling. Notice that since  the retarded interactions [Eq.~(\ref{eq:ret})] only appear in the diagonal parts (under the Fock basis) of the effective actions, they do not cause extra minus sign  problem.  Thus our QMC simulations can be considered as numerically exact since our ``system'' Hamiltonian (of hard-core bosons) is also positive definite.  Retarded interactions have also been implemented in other different QMC algorithms~\cite{Assaad2007, Weber2017}.


\begin{figure*}[htb]
\includegraphics[width=0.32\linewidth,bb=76 49 714 510]{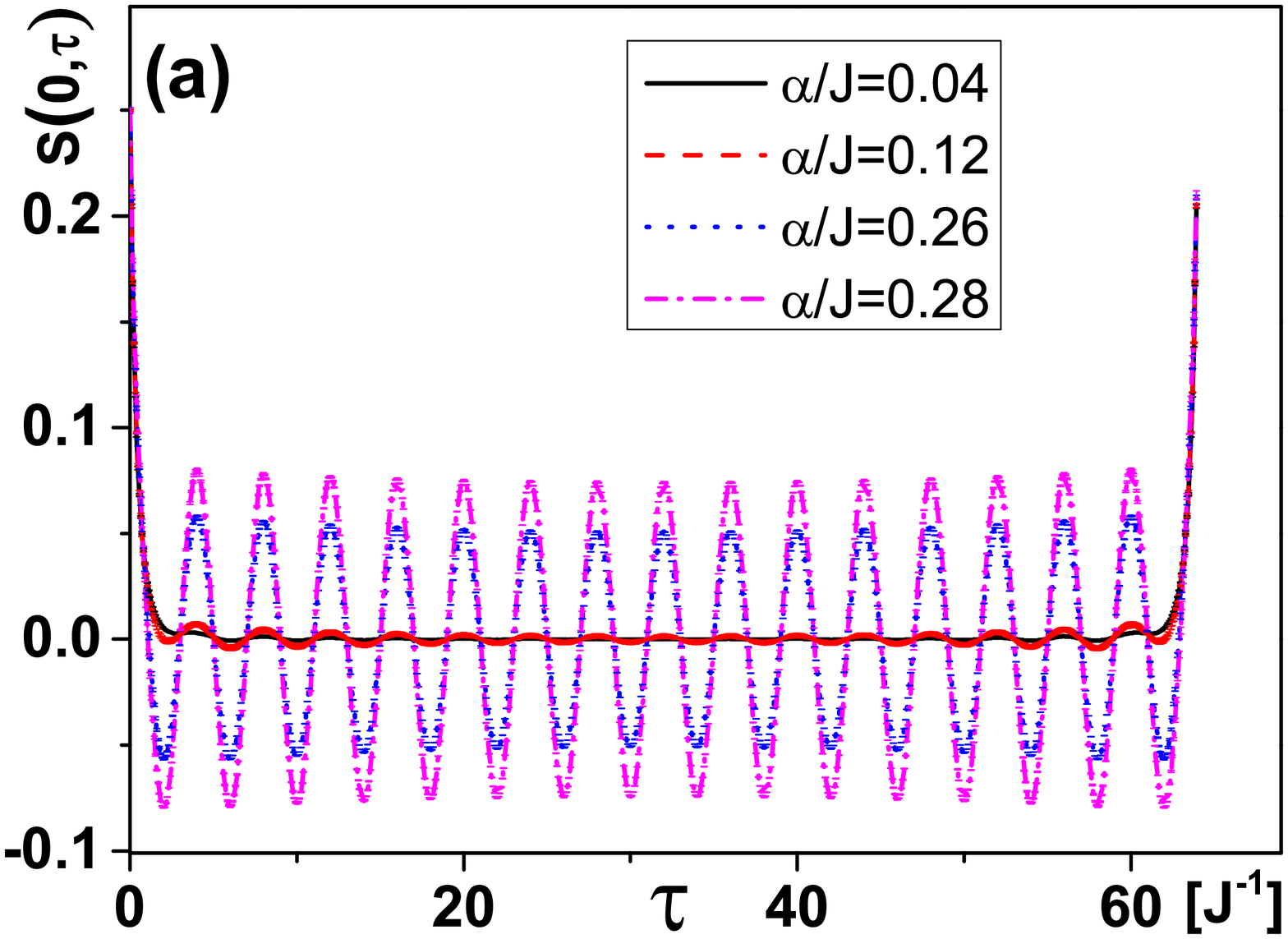}
\includegraphics[width=0.32\linewidth,bb=80 59 732 531]{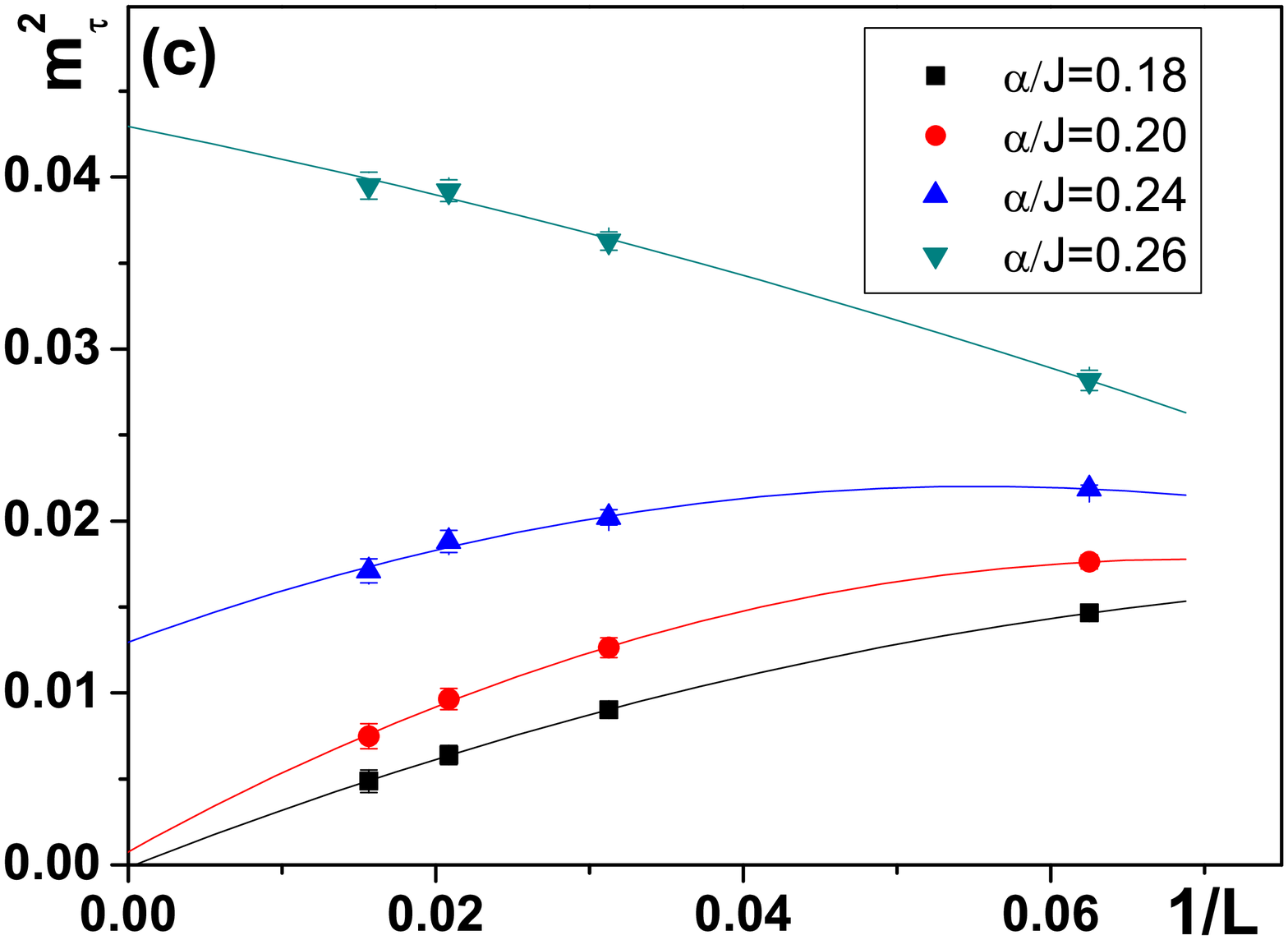}
\includegraphics[width=0.32\linewidth,bb=90 58 733 531]{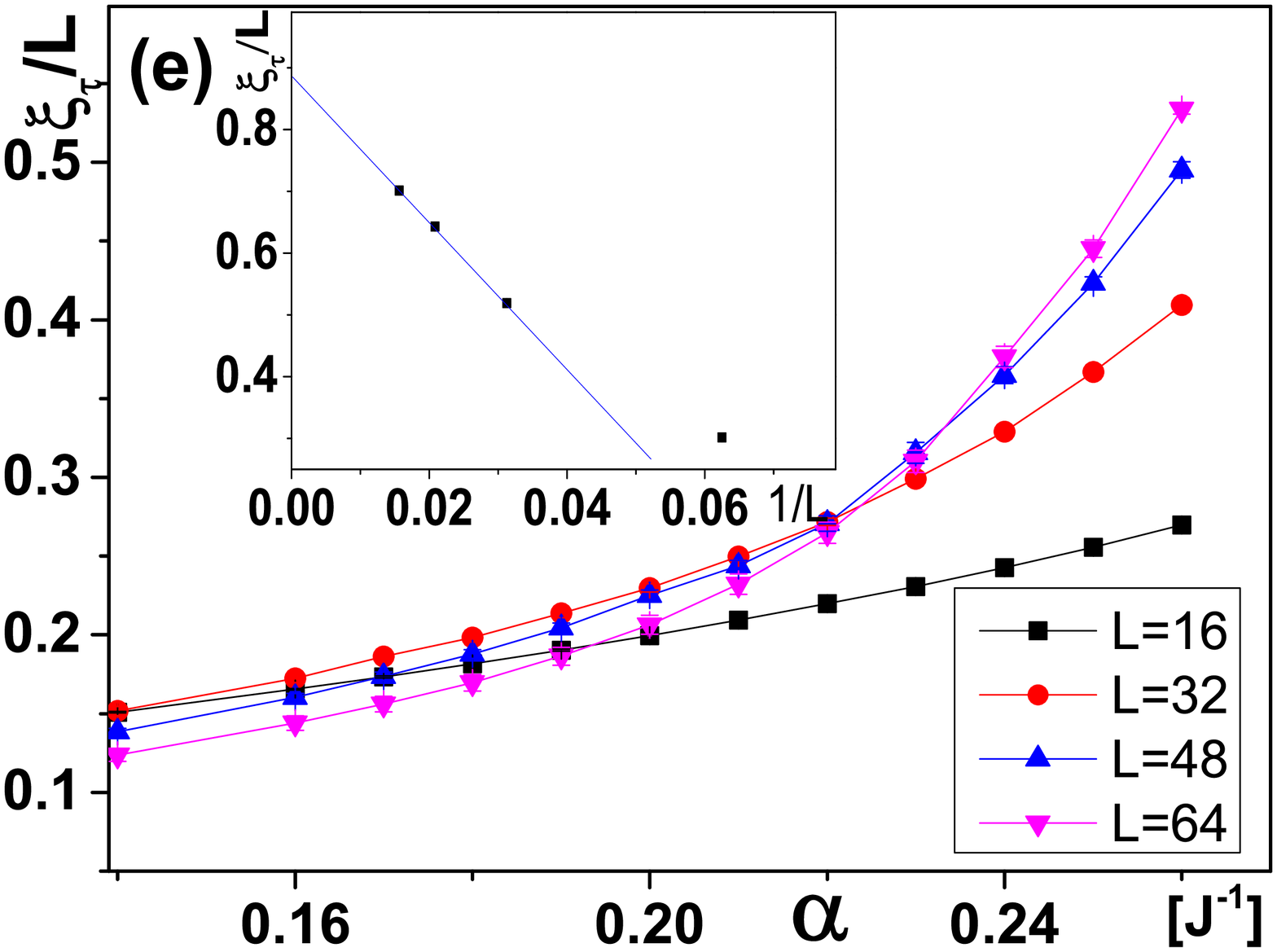}
\includegraphics[width=0.32\linewidth,bb=83 55 730 530]{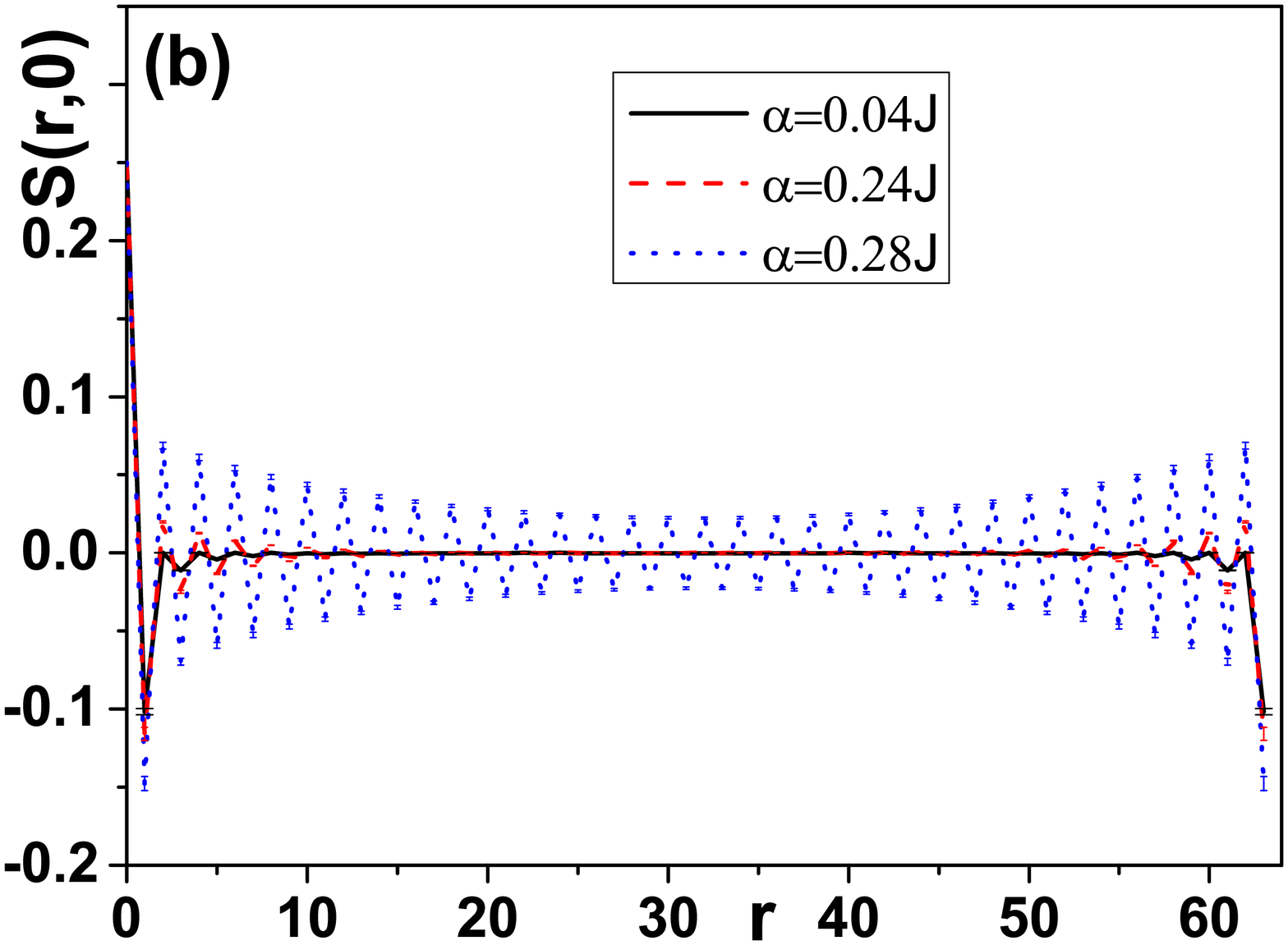}
\includegraphics[width=0.32\linewidth,bb=85 57 730 531]{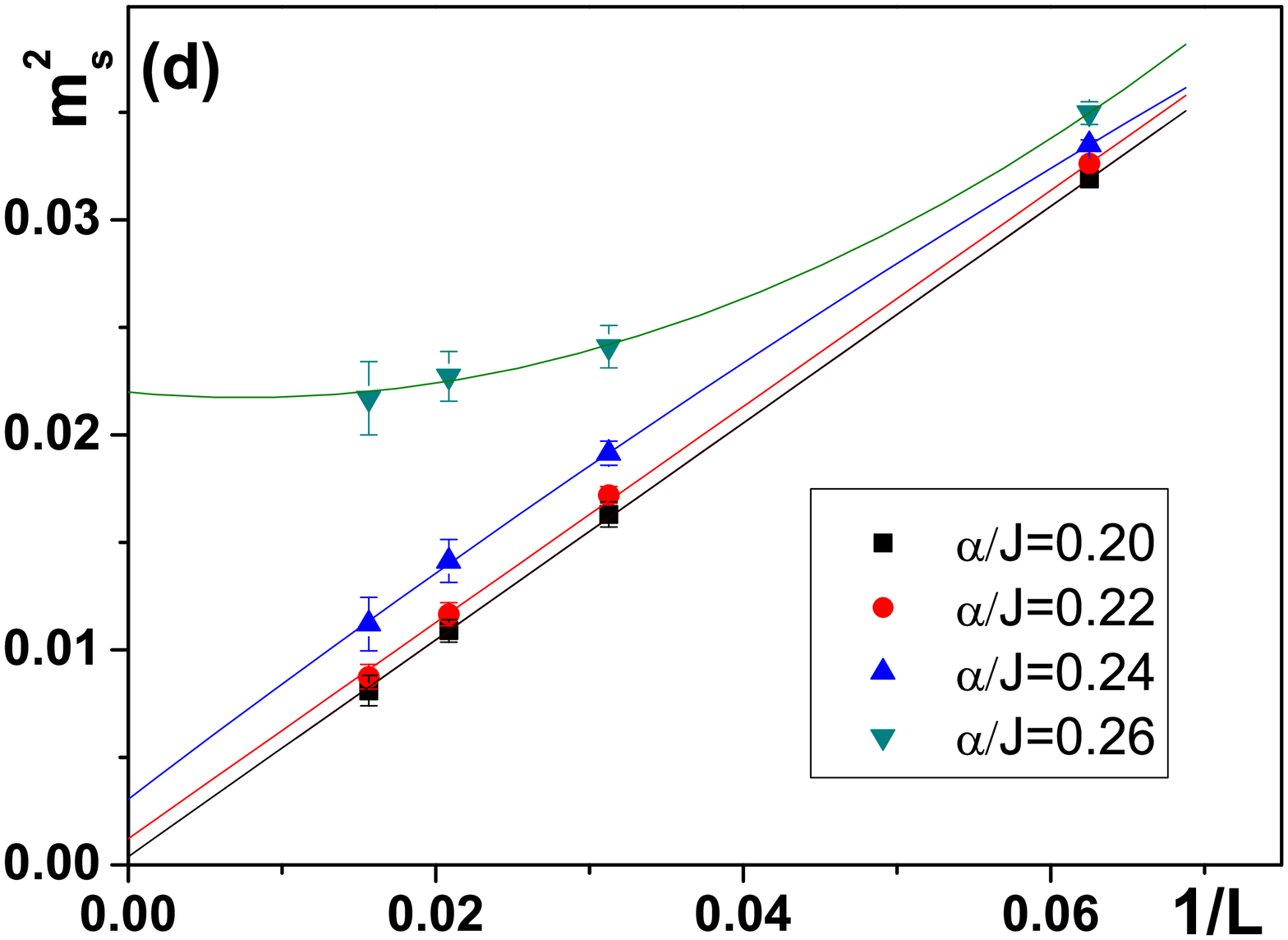}
\includegraphics[width=0.32\linewidth,bb=95 57 732 533]{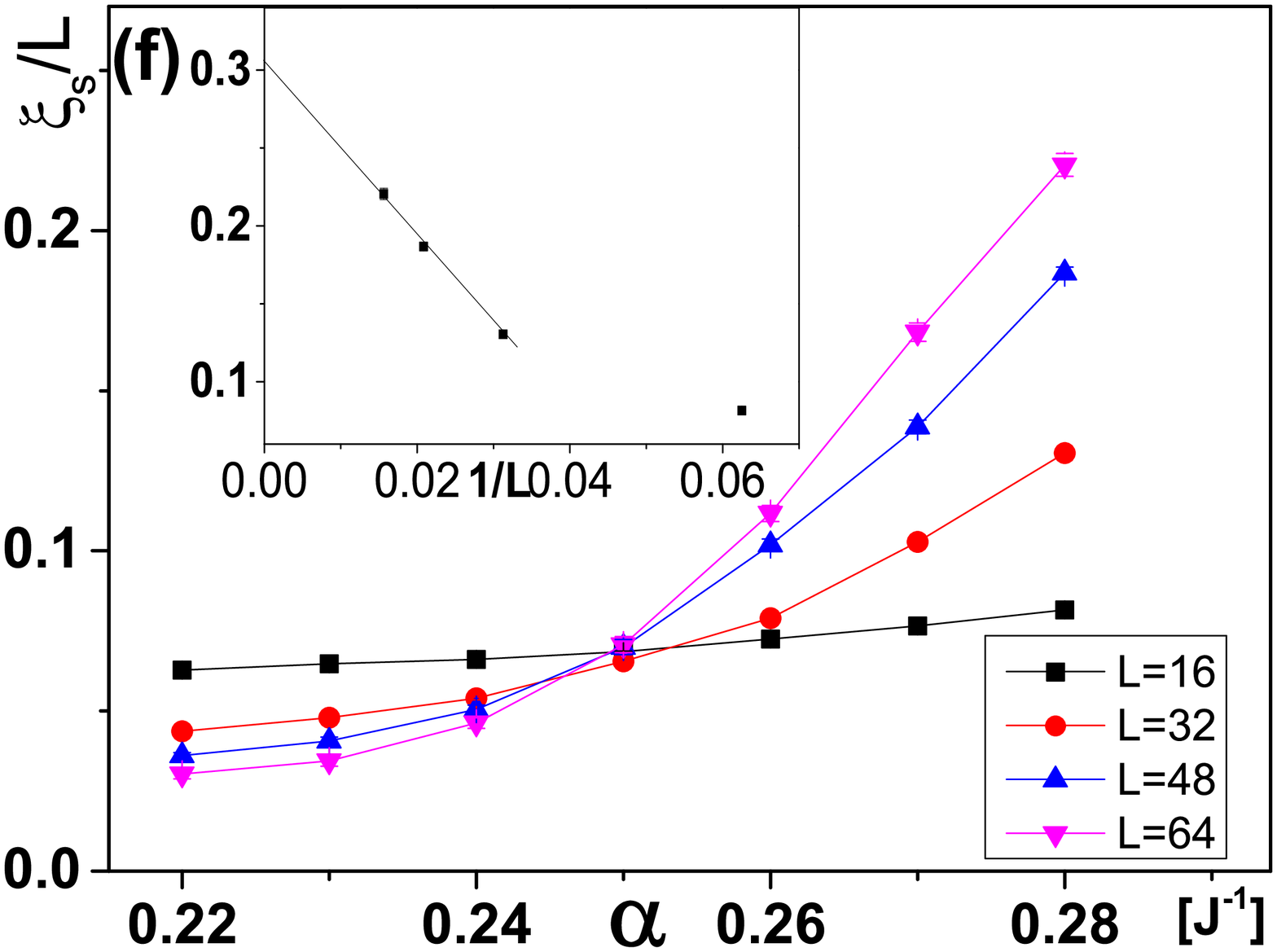}
\caption{(Color online). Temporal (upper panels) and spatial (lower panels) structures of iTC. (a) Onsite density correlation  as a function of $\tau$  and (b) equal-time density correlation as a function of $r$ for various retarded interaction strength $\alpha$ in a system with $L=64$; (c) and (d): Finite-size scaling of
 $m_\tau^2$  and  $m_s^2$  for different $\alpha$;  (e) normalized correlation time  and (f) normalized correlation length as a function of $\alpha$. Insets in (e) and (f): finite-size scaling of $\xi_\tau/\beta$ and $\xi_s/L$  with  $\alpha/J=0.28$.  For (a)-(f),  the scaling relation is $\beta=L$ and the parameters $\omega_c$ and $\omega_d$ is fixed as $\omega_c=J/4$ and $\omega_d=0.1J$.
} \label{fig:fig2}
\end{figure*}

\begin{figure*}[htb]
\includegraphics[width=0.32\linewidth,bb=90 57 733 532]{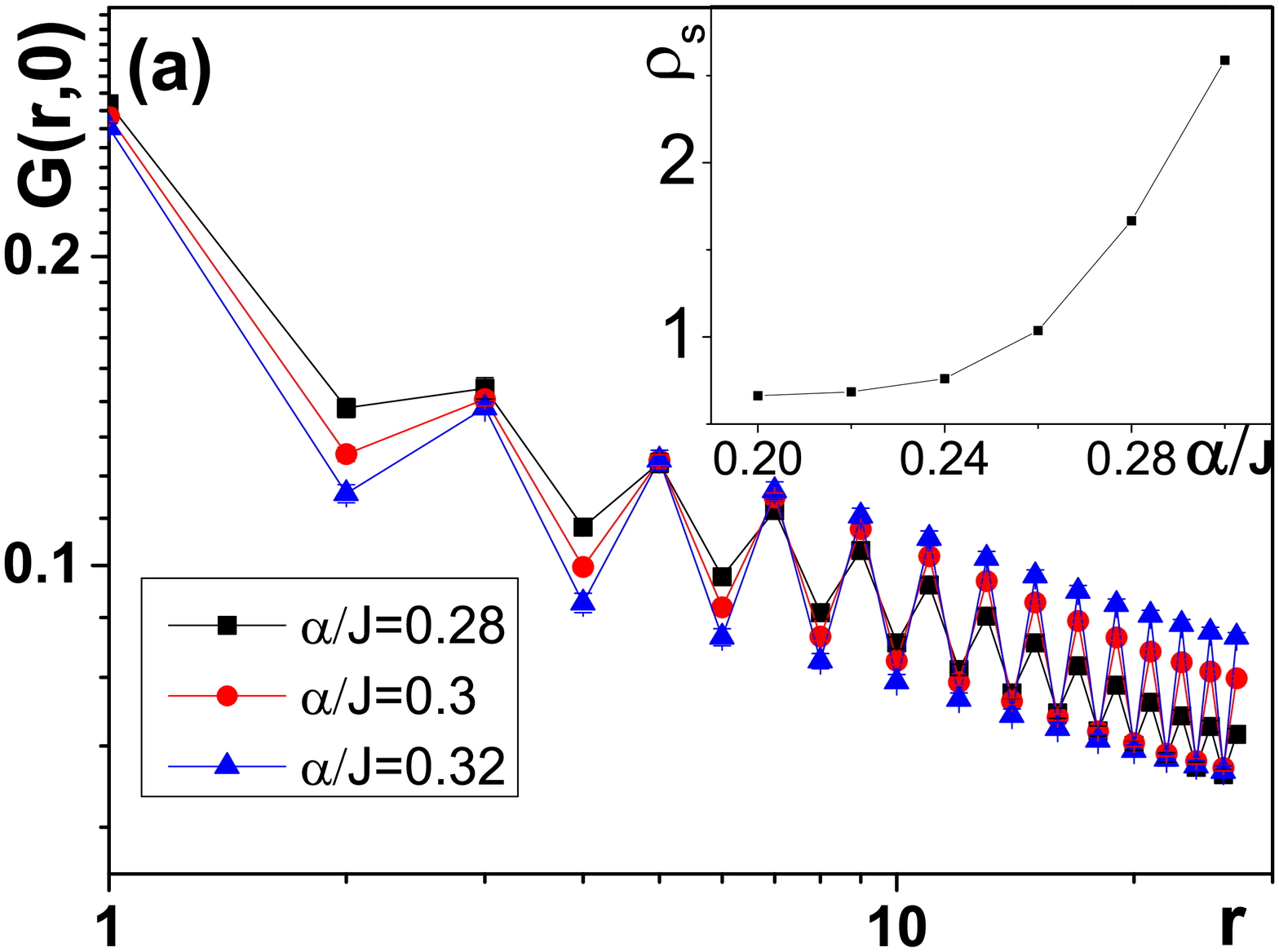}
\includegraphics[width=0.32\linewidth,bb=77 57 729 532]{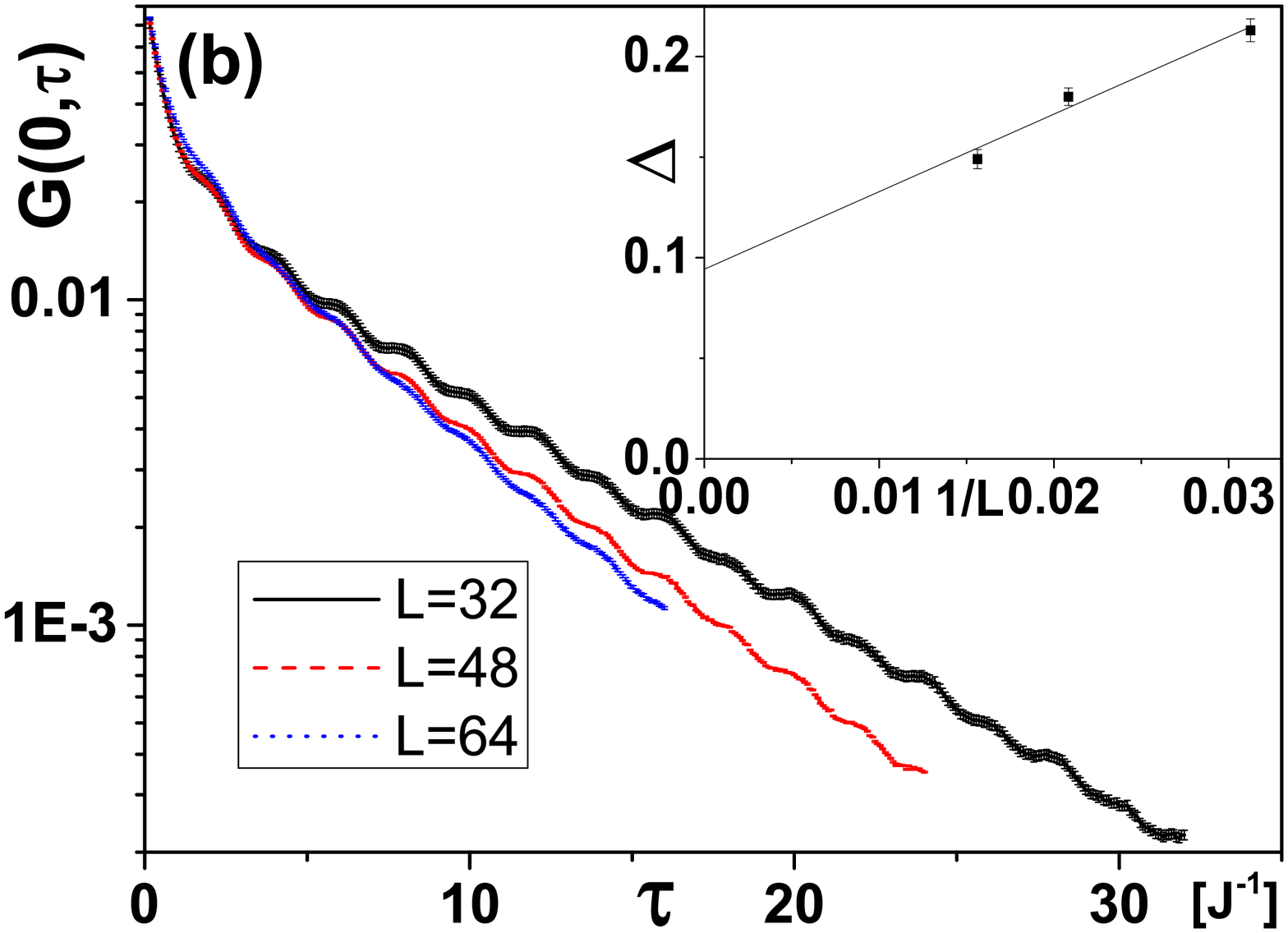}
\includegraphics[width=0.32\linewidth,bb=83 60 731 532]{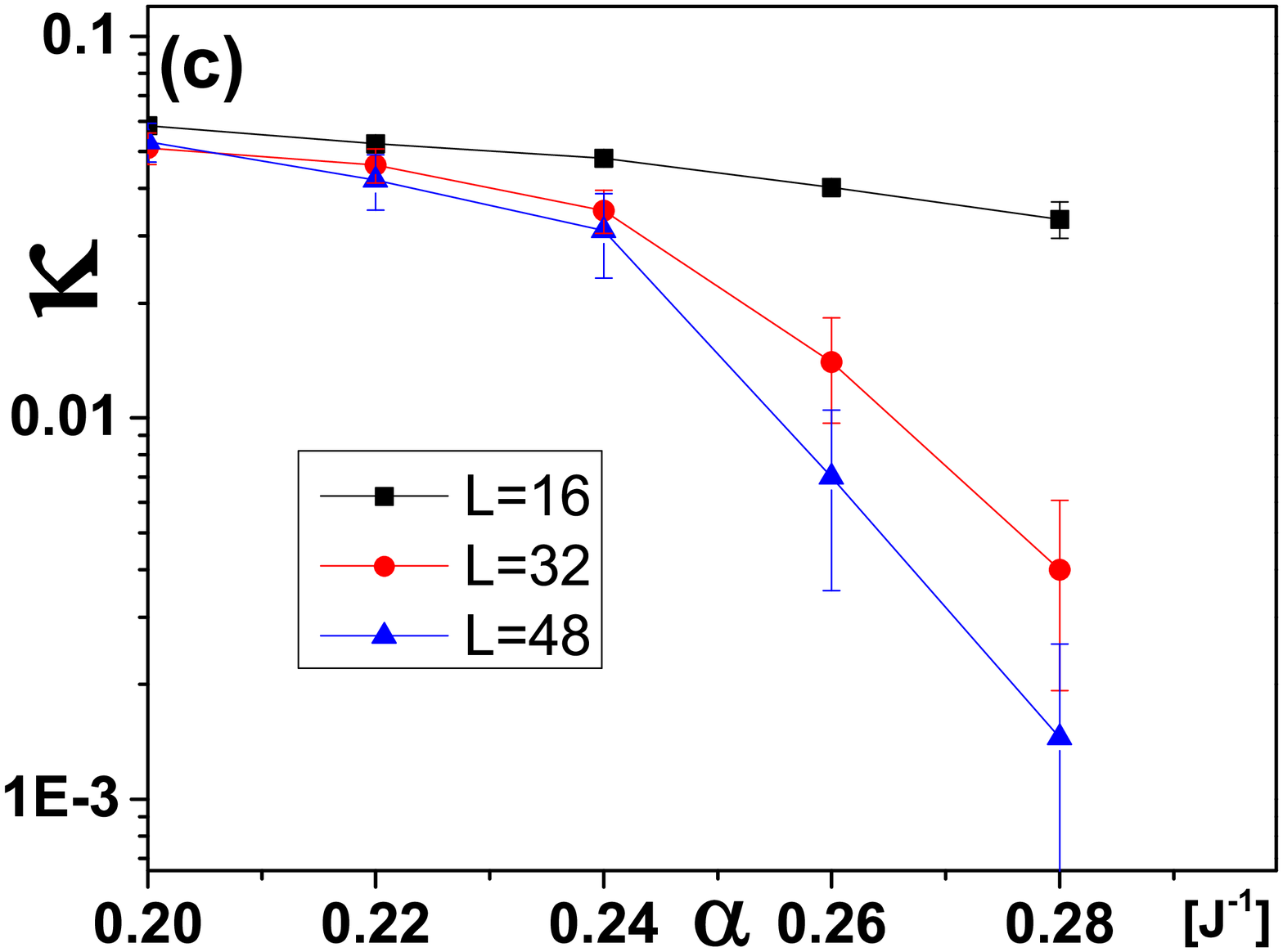}
\includegraphics[width=0.31\linewidth,bb=1 1 503 378]{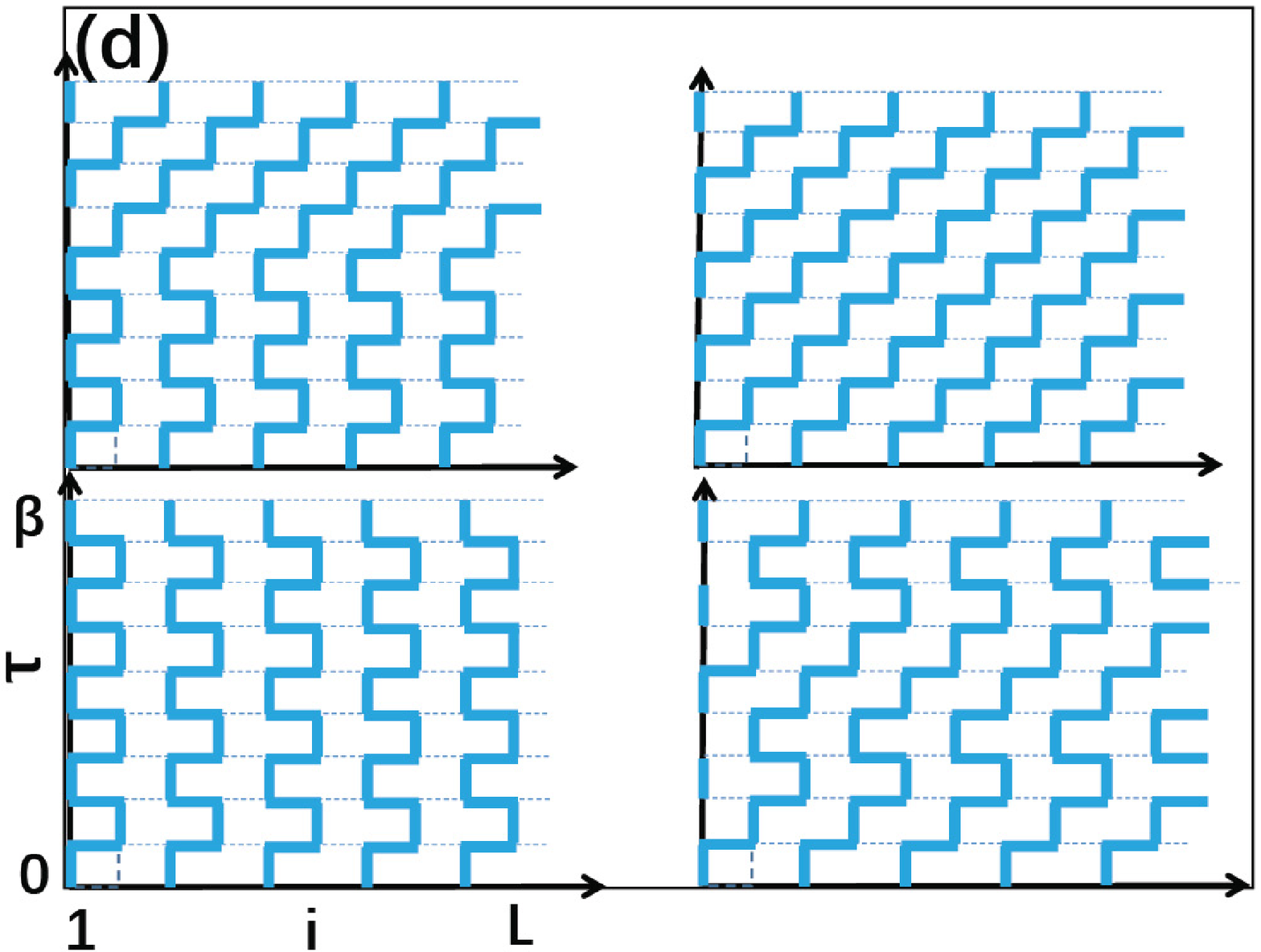}
\includegraphics[width=0.30\linewidth,bb=1 1 509 395]{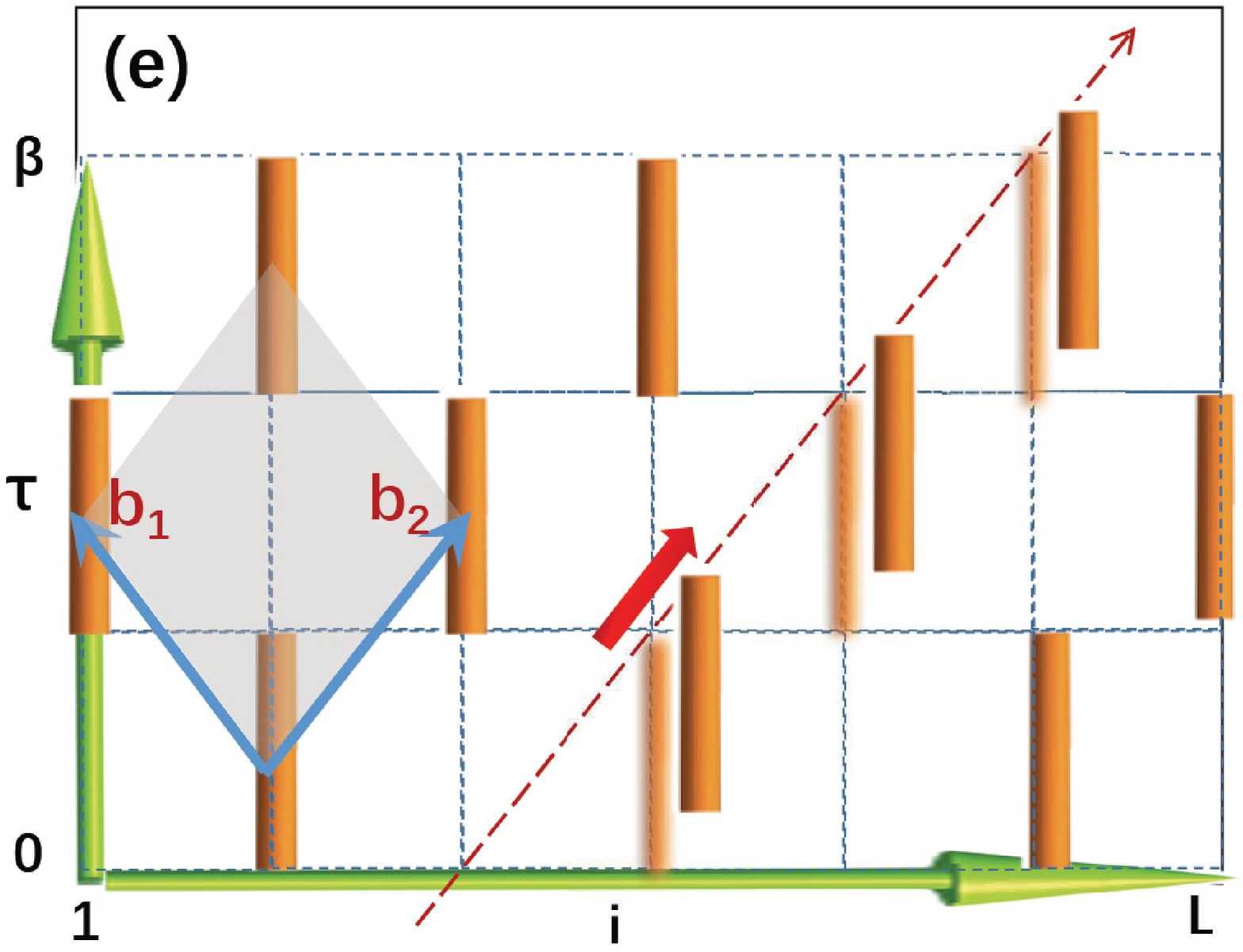}
\includegraphics[width=0.32\linewidth,bb=77 59 730 533]{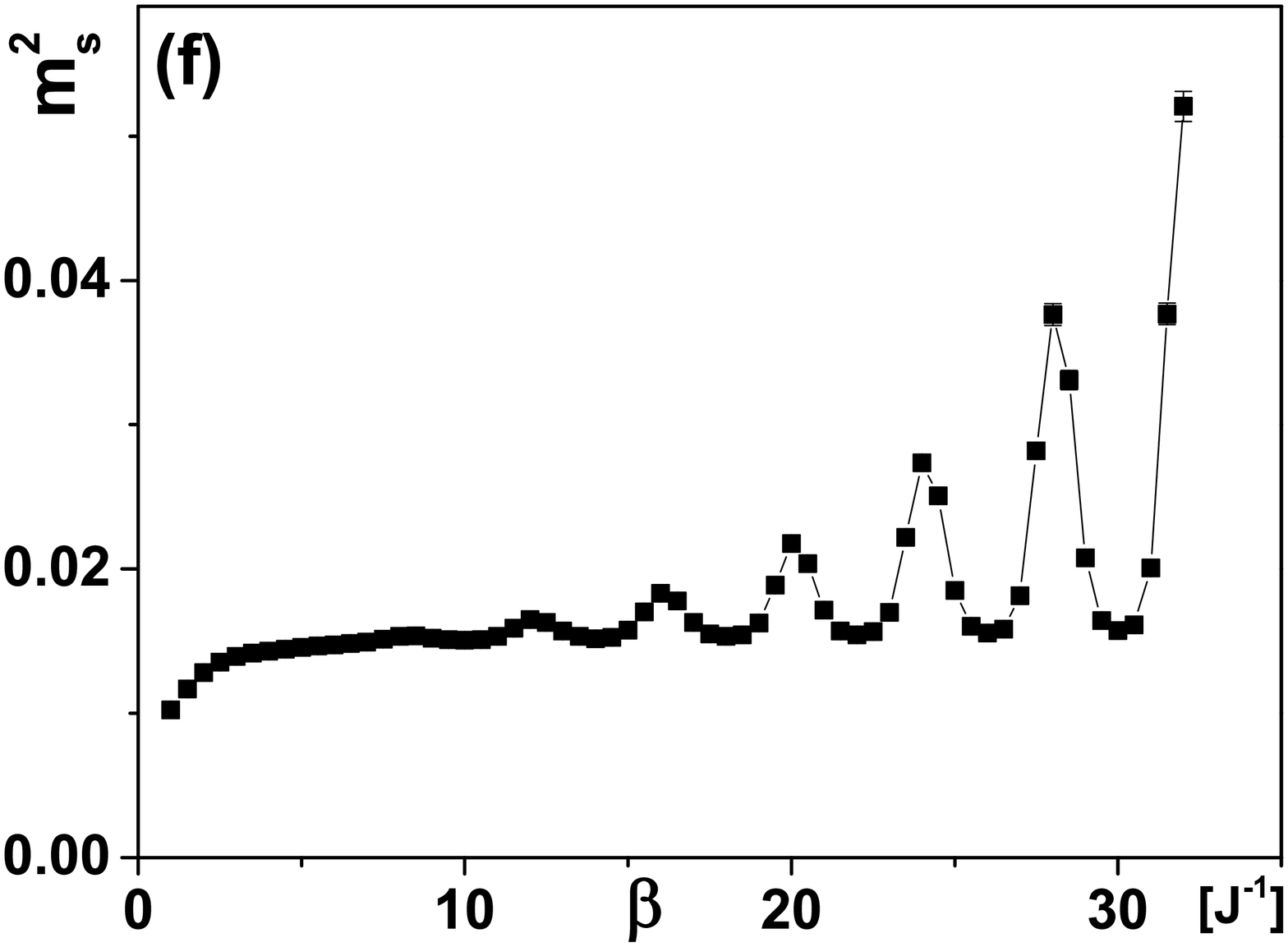}
\caption{(Color online).(a) Equal-time Green's function $G(r,0)$ for various $\alpha$ in the iTC phase (inset: The superfluid density as a function of $\alpha$)  with $L=64$. (b) Zero-momentum component of the unequal-time Green's  function $G(k=0,\tau)$ in the iTC phase and various system sizes (inset: the finite-size scaling of the exponent in the exponential decay of $G(0,\tau)\sim e^{-\Delta\tau}$). (c) The compressibility as a function of $\alpha$. (d) Typical temporal-spatial configurations that  equally contribute to the partition function in the strong coupling limit $J/\alpha\rightarrow 0$. (e) Sketch of temporal-spatial crystalline structure of the iTC phase and its unit cell (the grey rhombic area spanned by primitive basis vectors $\mathbf{b}_{1/2}$). (f) The (inverse) temperature dependence of $m_s^2$ in the  iTC phase  and system size $L=32$. $L=\beta$ for (a)-(c) and $\alpha=0.3J$ for (b) and (f).  $\omega_c$ and $\omega_d$ are the same as in Fig.\ref{fig:fig1}.
} \label{fig:fig3}
\end{figure*}

\section{\bf Groundstate properties.}
The bosonic models with retarded interactions can be solved numerical exactly by Quantum Monte Carlo simulations (see Methods).   We first focus on the ground state  of  the total system (system+bath). In finite size scaling,  the inverse temperature is chosen as $\beta=L(2L)$ for 1D (square) lattice, and the ground state in the thermodynamic limit is approached in the limit $L\rightarrow \infty$. In the 1D case without retardation,  the ground state is a quasi-superfluid ground state characterized by algebraic decaying correlation functions in both space and iTime.  As the retarded interaction is switched on, spatial-temporal configurations with periodically oscillating world lines are favored since the interacting energies are lowered in these patterns. The competition between the hopping and the retarded interactions will lead to novel quantum phases.

The crystalline order can be identified from the density correlation functions in space and iTime:
$ S(r,\tau)=\frac 1L\sum_i \langle (n_i(0)-n_0)(n_{i+r}(\tau)-n_0)\rangle$.
We first focus on the onsite correlations with $r=0$. As shown in Fig.~\ref{fig:fig2}(a), for small $\alpha$, $S(0,\tau)$ still decays algebraically with $\tau$, but is accompanied by oscillations. While for sufficiently large $\alpha$ a long-range order is built up and a characteristic time period emerges, indicating a spontaneous breaking of translational symmetry in iTime,  characterized by the order parameter $m^2_\tau=\frac 1\beta \int_0^\beta d\tau S(0,\tau) \cos2\pi\omega_c\tau.$
 A finite-size scaling of $m_\tau$ is shown in Fig.\ref{fig:fig2}(c), from which we can find that for large $\alpha$,  $m_\tau$ extrapolates to a finite value in the thermodynamic limit, indicating a crystalline order in iTime. The existence of a quantum phase transition can be further verified by the correlation length $\xi_\tau$ along iTime ($S(0,\tau)\sim e^{-|\tau|/\xi_\tau}$ for sufficiently large $\tau$),  which can be derived from the structure factor (Fourier transformation of $S(0,\tau)$)~\cite{Sandvik2010}. The  normalized correlation length $\xi_\tau/\beta$  as a function of $\alpha$ for different system size and $\beta$ is plotted in  Fig.~\ref{fig:fig2}(e), where we find a crossing point at $\alpha_c^\tau=0.23(1)$, indicating a scaling invariant critical point. In the iTC phase, $\xi_\tau/\beta$ linearly scales with the system size, and extrapolates to a finite value in the thermodynamic limit (inset of Fig.~\ref{fig:fig2}(e)).

 For sufficiently large $\alpha$, it is interesting to notice that the crystalline pattern not only emerges in iTime, but also in real space, which can be identified by the equal-time correlation function $S(r,0)$ as well as its structure factor $S(Q)=\frac 1L\sum_r e^{iQr}S(r,0)$. In the case of half-filling, the spatial crystalline pattern is a charge-density-wave (CDW) state with a periodicity twice that of the lattice, as shown in Fig.~\ref{fig:fig2}(b). A finite size scaling of the order parameter $m_s=\sqrt{S(Q=\pi)}$ is shown in Fig.~\ref{fig:fig2}(d). Since there is no direct nearest-neighboring repulsive interaction in our system Hamiltonian, the CDW order is formed by an effective repulsive interaction induced by the retardation: the retarded interaction~(\ref{eq:interaction}) favors oscillating world lines in iTime, and thus requires the hard-core bosons being separated as far as possible to avoid blocking the oscillations of each other.  Therefore at half-filling the tendency towards CDW order is expected in real space. Similar to the analysis above, we also plot the normalized correlation length $\xi_s/L$ in real space  as function of $\alpha$ in Fig.\ref{fig:fig2}(f), where a continuous phase transition is found at $\alpha=0.245(5)$.  Above this value, long-range CDW correlations emerge in real space.


To further investigate the properties of the iTC, we calculate the single particle Green's  function: $G(r,\tau)=\langle a_i^\dag(\tau) a_{i+r}(0)\rangle$. The equal-time part $G(r,0)$ is shown in Fig.~\ref{fig:fig3}(a),  which decays  algebraically with $r$ (accompanied by an oscillation) even for large $\alpha$, indicating  a quasi-superfluid order in the iTC. This order can be further verified from the superfluid density, which can be calculated as $\rho_s=\frac L\beta\langle W^2\rangle$, where $W$ is the winding number denoting  the net times by which the world-lines wrap around the 1D
lattice.  As shown in the inset of Fig.~\ref{fig:fig3}(a), $\rho_s$ grows monotonically with $\alpha$, indicating that the retarded interactions enhance the mobility of bosons. Now we turn to the temporal part of $G(r,\tau)$, and focus on its $k=0$ component: $G(k=0,\tau)=\frac 1L \sum_r G(r,\tau)$. Fig.~\ref{fig:fig3}(b) shows that $G(k=0,\tau)$ decays exponentially with $\tau$, indicating a finite single-particle charge gap. This agrees with the results of the compressibility  $\kappa=\frac\beta L(\langle N^2\rangle-\langle N\rangle^2)$ with $N$ the total particle number.  Fig.~\ref{fig:fig3}(c) show that $\kappa$ decays to zero for large $\alpha$, thus the iTC is incompressible.

 In summary, the iTC phase induced by the retarded interaction~(\ref{eq:interaction}) spontaneously breaks the translational symmetries in both real space and iTime. Besides these intriguing properties, this phase is unique in the sense that it is incompressible but  possess quasi-long range superfluid order, so it is fundamentally different from the conventional bosonic ground states in any closed quantum many-body systems: e.g. the (quasi-)superfluidity, Mott insulator, supersolid, Bose glass or Bose metal. The nature of this phase can be qualitatively understood in the strong coupling limit $\alpha\gg J$, where the retarded interaction dominates in the partition function.  For each site, the retarded interaction~(\ref{eq:interaction}) favors configurations with alternating $n(\tau)$ that oscillates between 0 and 1 with a frequency $2\pi\omega_c$, which  is reminiscent  of the soliton solutions in the BCS pairing model~\cite{Mukhin2009,Galitski2010},  similar states have recently been proved to be metastable in a wide class of closed quantum many-body systems~\cite{Mukhin2019}.    Since a world line needs to be continuous, the temporal-spatial configurations  minimizing the retarded interacting action~(\ref{eq:ret}) are highly degenerate. As shown in the Fig.~\ref{fig:fig3}(d), these configurations can be understood as a synchronous movement of the bosons in iTime, in the sense that different bosons hop simultaneously towards the same direction; These synchronous movements can reduce the possibility of the collisions of world lines of different hard-core bosons, and at the same time, minimize the retarded interacting energy.    In this hugely degenerate manifold, those configurations with nonzero winding number contribute to the nonzero superfluid density (e.g. the 2nd figure in Fig.\ref{fig:fig3}(d)). The incompressible nature can also be understood in such a picture: an extra particle added into the half-filled system will inevitably block or impede the synchronous movement of other bosons due to its hard-core nature, and thus will increase the retarded interacting energy and give rise to a finite charge gap.

{\it Symmetries and excitations.} The iTC phase simultaneously breaks the continuous and discrete translational symmetry in temporal and spatial directions.  The spacial and temporal spontaneous symmetry breakings lock with each other,  leaving  an unbroken subgroup of  intertwined space-iTime translational symmetry. The unit cell of the iTC is spanned by the primitive basis vectors $\mathbf{b}_{1/2}$ with a rhombic symmetry as shown in Fig.~\ref{fig:fig3}(e),  as opposed to a ``decoupled'' direct product of vectors in $\tau$ and $x$ directions with a rectangular symmetry.
Due to the fact that the time continuum is locked with the discrete spatial lattice we started with,  small translational fluctuations of the atomic positions along the primitive basis vectors  do not generate gapless ``phonons'', but  excitations with finite energy cost due to the discreteness of the spatial translational symmetry in our model. The space-iTime locking avoids  the dangerous gapless modes in renormalization, so remarkably stabilizes  the iTC phase. 

\section{Temperature effect of iTime crystal.}
 The most striking effect of  iTC phase can be found at finite temperature.  In conventional systems, the temperature dependence of a physical quantity could be either monotonic or non-monotonic with one or several peaks. In contrast,  an iTC could exhibit exotic thermal behavior that its physical observable depends on temperature in an oscillating way. This intriguing property can be understood as follows: the crystal structure in iTime will perfectly fit at certain temperatures satisfying the condition that $\beta$ is a multiple of the period of iTC; otherwise, for general $\beta$ the mismatch between them will introduce defects or distortions, which impede the synchronous movement of the bosons, thus are detrimental to crystalline order in both iTime and space.  As an example, in Fig.~\ref{fig:fig3}(f), we plot the CDW ordering parameter as a function of $\beta$, which shows an oscillating behavior with peaks at the positions of integer multiples of the iTC period. This abnormal  thermal behavior can be considered as an experimental diagnostic of the iTC phase.

\begin{figure}[htb]
\includegraphics[width=0.98\linewidth,bb=69 42 914 522]{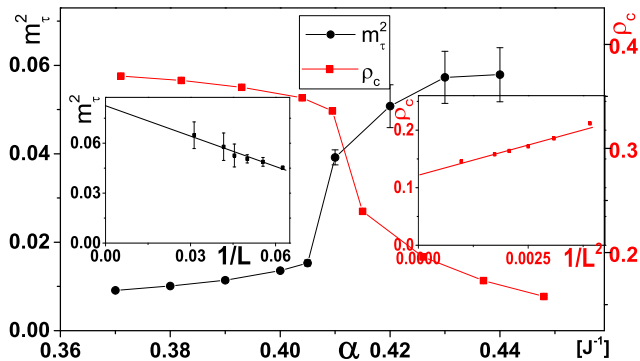}
\caption{(Color online)The iTC order parameter $m_\tau^2$ (black circle) and condensate fraction $\rho_c$ (red square) as functions of $\alpha$ in a $L\times L$ square lattice model with $2L=\beta=48$. Inset: the finite-size scaling of $m_\tau^2$ (left) and $\rho_c$ (right) in the 2D iTC phase with $\alpha=0.44J$, $\beta=2L$, $\omega_c=J/4$ and $\omega_d=0.1J$.
} \label{fig:fig4}
\end{figure}

{\it Extension to square lattices.}  Up to now, we have focused on 1D in the spatial part, while a generalization of our method to  2D square lattices is straightforward. Similar to the 1D case, we find that the temporal crystalline order also emerges for sufficiently large retarded interactions (left inset of Fig.~\ref{fig:fig4}).  For  a fixed $L$ and $\beta$ there seems to be a jump of the iTC order parameters $m_\tau^2$ as a function of $\alpha$ (Fig.~\ref{fig:fig4}), indicating that it is likely a 1st order quantum phase transition. However, due to the finite size effect and the limited system size in our simulation, we cannot preclude other possibilities.
Another important difference between 1D and 2D is that in the latter case, there exists true superfluidity long-range order characterized by the condensate fraction $\rho_c=\frac{1}{L^2}\sum_r G(r,0)$. In Fig.~\ref{fig:fig4} (b), we plot $\rho_c$ as a function of $\alpha$, where we also find a discontinuity occurs at $\alpha_c=0.405(5) J$. We find that even in the iTC phase, the superfluid order still persists (right inset of Fig.~\ref{fig:fig4}).

\section{\bf Conclusion and outlook}
We introduce a class of quantum many-body systems with oscillating retarded interactions in iTime and performed numerically exact QMC simulations. Our study finds an iTC phase with exotic zero-temperature properties and abnormal thermal behavior. To conclude this paper, we wish to outline some differences between our results and other related studies recently. {\it First,} our model is conveniently presented by a Euclidean action, so it is beyond the scope of Hamiltonian problems and does not contradict the ``no time crystal ground state'' theorem~\cite{Bruno2013,Watanabe2015}.
{\it Second,} even though it corresponds to a ``time''-dependent problem, the ``time'' dependence enters our model through the relative time difference in terms of two-body interactions, in contrast to the early Floquet models whose time-dependence  is explicitly through overall time ~\cite{Else2016,Yao2017}. This difference fundamentally changes the symmetry and a continuous symmetry model opens the wonder of what the analogue of Goldstone modes in spatial crystals, i.e,  the ubiquitous  and all important phonons, will be after the spontaneous breakdown of time translational symmetry.   The potential of discovering ``time'' phonons is now open for further work.
{\it Finally,} a thermodynamic physical system is found to support our  ``theoretical'' time-crystal  model.     Standard statistical physics is applied to show the correspondence of our imaginary-time quantum mechanical model with retarded interaction (dependent on relative time!) to the ensemble of hard-core bosons embedded in engineered baths. This report therefore extends  the study of time crystal to the Euclidean spacetime corresponding to thermal ensemble systems.

Our model can be derived from open quantum systems that have been studied extensively, such as quantum optomechanical resonators, QED cavities of interacting light and atoms, trapped ions~\cite{Barreiro2011} and Rydberg atoms~\cite{Bernien2017}) coupled to synthetic thermal baths, etc. The crucial ingredient to realize the proposed idea  is to look for a system that has retarded interaction and allows the system-bath coupling to be easily tuned. The above main conclusions are firmly supported by numerical exact QMC simulations of bosonic models with oscillating retarded interactions in relative iTime.  Our method can be  straightforwardly generalized to other bosonic and spin systems with retarded interactions, in one or higher dimensions. The competition between time-oscillating interaction and quantum-thermal fluctuations is expected to give rise to other forms of nontrivial temporal-thermal orderings that are  completely absent in the conventional Hamiltonian systems.   A generalization to fermionic models may be even more interesting since it opens up new possibilities of finding novel ``non-fermi liquid''.
It is also interesting to study the real time counterpart of our model, which may be derived by integrating out a non-equilibrium bath.  This may provide a new perspective for current time crystal researches since such an effective action  has the full continuous (rather than discrete) symmetry in temporal translation.

\noindent{\em Acknowledgment.} We appreciate insightful discussions with F. Wilczek and V. Galitski.  This work is supported in part by the National Key Research and Development Program of China (Grant No. 2016YFA0302001), NSFC of  China (Grant No. 11674221 and No.11745006), Shanghai Rising-Star Program, Eastern Scholar Professor of Distinguished Appointment Program and Project of Thousand Youth Talents (Z. C.) and by AFOSR Grant No.  FA9550-16-1-0006,  MURI-ARO Grant No. W911NF-17-1-0323 through UC Santa Barbara,  and NSF China Overseas Scholar Collaborative Program Grant No. 11429402 sponsored by Peking University (W.V. L.).

{\it Note:} During the completion of this manuscript, we notice the preprint by K.B. Efetov~\cite{Efetov2019} on thermodynamic time crystal. The subject is related but the result and models are completely different and independent.

\appendix

\section{A microscopic model of fermionic bath}
\label{sec:bath}
In this section, we provide a microscopic model of the bath that can induce the
retarded interaction as shown in the main text. The bath is composed of
fermionic quasiparticles in interacting or disordered quantum many-body
systems. Recently, Kozii and Fu ~\cite{Kozii2017} proposed an effective
non-Hermitian Hamiltonian to described such quasiparticles, where the effect of
interaction or disorder can be encoded into the quasiparticle's self-energy,
whose imaginary part represents a finite lifetime of the quasiparticles. For a
quasiparticle with a given momentum, its lifetime can be determined by the
electron-electron (or electron-phonon) scattering, or the scattering by impurities.  Even
though a complete microscopic Hamiltonian of an interacting fermionic system
should be Hermitian, the finite lifetime of the quasiparticle give rise to the
non-Hermitianity of the effective Hamiltonian description of the quasiparticles.

Motivated by Ref.~\cite{Kozii2017},  we propose a microscopic model for the bath
of quasi-particles, which can be effectively
described by a two-band fermionic non-Hermitian Hamiltonian as: $H=\sum_i
\mathbf{\Psi}_i^\dag \hat{H}_i \mathbf{\Psi}_i$, where
$\mathbf{\Psi}_i=[\Psi_{i,1},\Psi_{i,2}]^T$, and $\Psi_{i,1(2)}$ is an
annihilating operator of the bath quasiparticle at site $i$ with band index
$1(2)$, and
\begin{equation}
\hat{H}_i= \left(
  \begin{array}{cc}
    \Sigma_1(k,\omega) & \omega_d  \\
    \omega_d & \Sigma_2(k,\omega) \\
\end{array}
\right) \,, \label{M0}
\end{equation}
where $\omega_d$ is interband hopping amplitude, and we assume that the bath at each lattice site is independent of
others. $\Sigma_{1(2)}(k,\omega)$ is the bath quasi-particle self-energy,
whose real and imaginary part correspond to its dispersion relation and decay
rate, respectively. Since the bath under consideration here  is spatially local, thus we can neglect
its dispersion and only keep its imaginary part. For simplicity, we only keep
the constant terms in the imaginary part, thus the self energy can be
approximated as $\Sigma_1(k,\omega)=\Sigma_2(k,\omega)\approx -i\Gamma$. In
Ref.\cite{Kozii2017}, a microscopic derivation of $\Gamma$ from an
electron-phonon interacting model has been discussed.

By performing the standard procedure of path integral for
fermions~\cite{Nagaosa1999}, the partition function can be expressed as $Z=\int
\mathfrak{D}[\bar{\psi}]\mathfrak{D}[\psi]e^{-S[\bar{\psi},\psi]}$, where
$\psi_i$ is the Grassmann number satisfying
$\Psi|\psi\rangle=\psi|\psi\rangle$ with $|\psi\rangle$ the fermonic coherent
state (we neglect the site and band indexes here for simplicity), and $\mathfrak{D}[\psi]$ denotes the functional integral over $\psi(\tau)$. The effective action of the bath can be written as:
\begin{eqnarray}
S&=&\int_0^\beta d\tau  [\bar{\psi}(\tau) \partial_\tau \psi(\tau)-\bar{\psi}(\tau) \hat{H} \psi(\tau)]\\
&=&\sum_m \bar{\psi}(-i\omega_m)[-i\omega_m\hat{1}-\hat{H}]\psi(i\omega_m) ,
\end{eqnarray}
where we performed the Fourier transformation: $\psi(\tau)=\frac 1\beta\sum_m
e^{-i\omega_m\tau} \psi(i\omega_m)$; $\bar{\psi}(\tau)=\frac 1\beta\sum_m
e^{i\omega_m\tau} \psi(i\omega_m)$, and $\omega_m=\pi T (2m+1)$ is the Matsubara
frequency.

\begin{figure*}[htb]
\includegraphics[width=0.32\linewidth,bb=0 0 295 212]{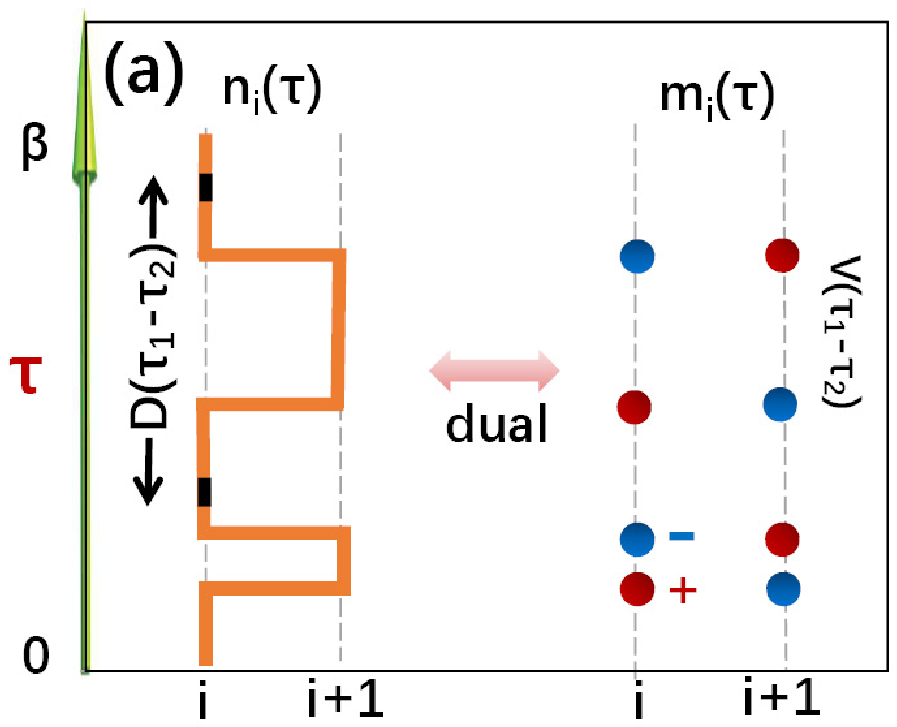}
\includegraphics[width=0.32\linewidth,bb=58 45 680 488]{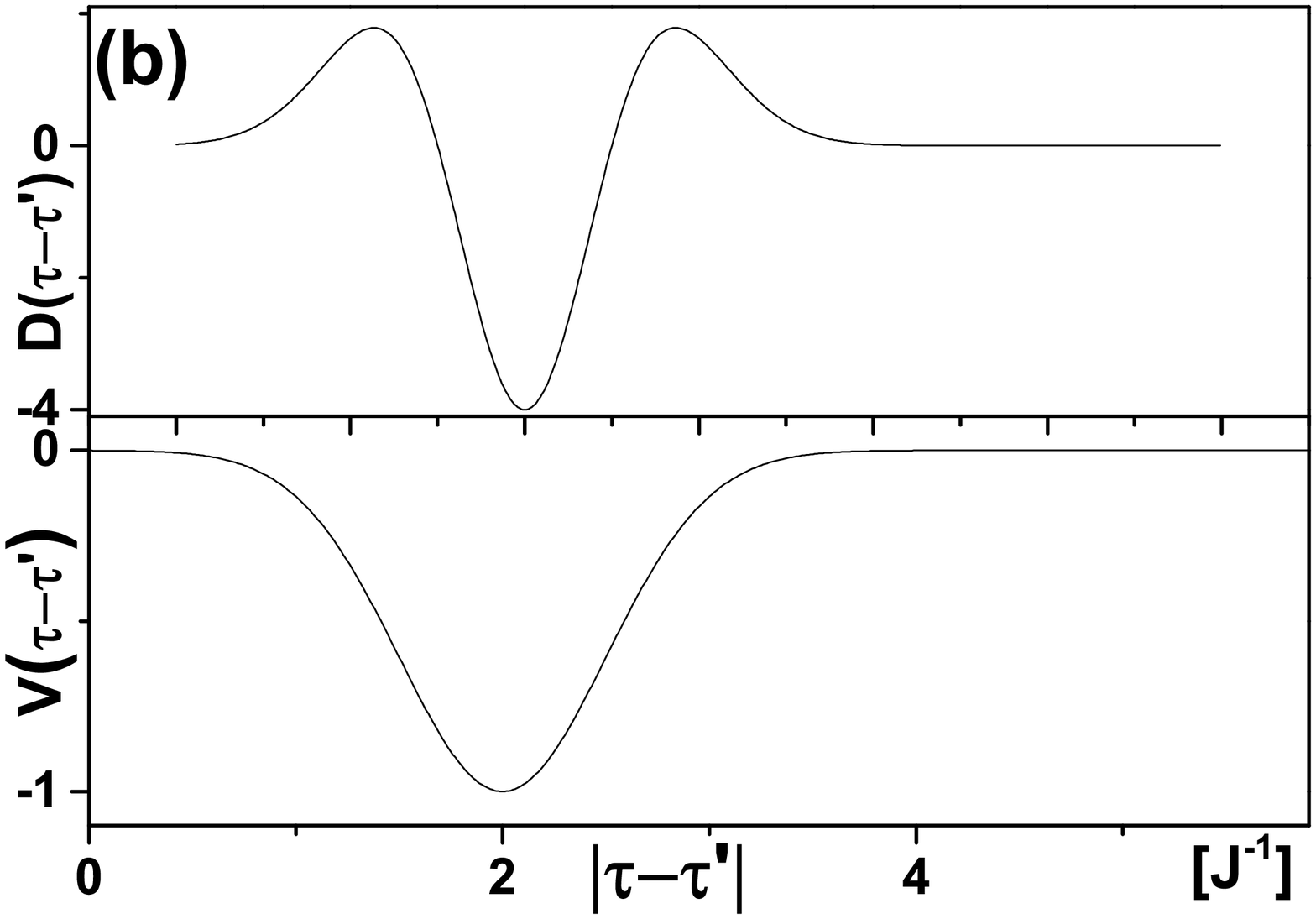}
\includegraphics[width=0.335\linewidth,bb=50 58 733 531]{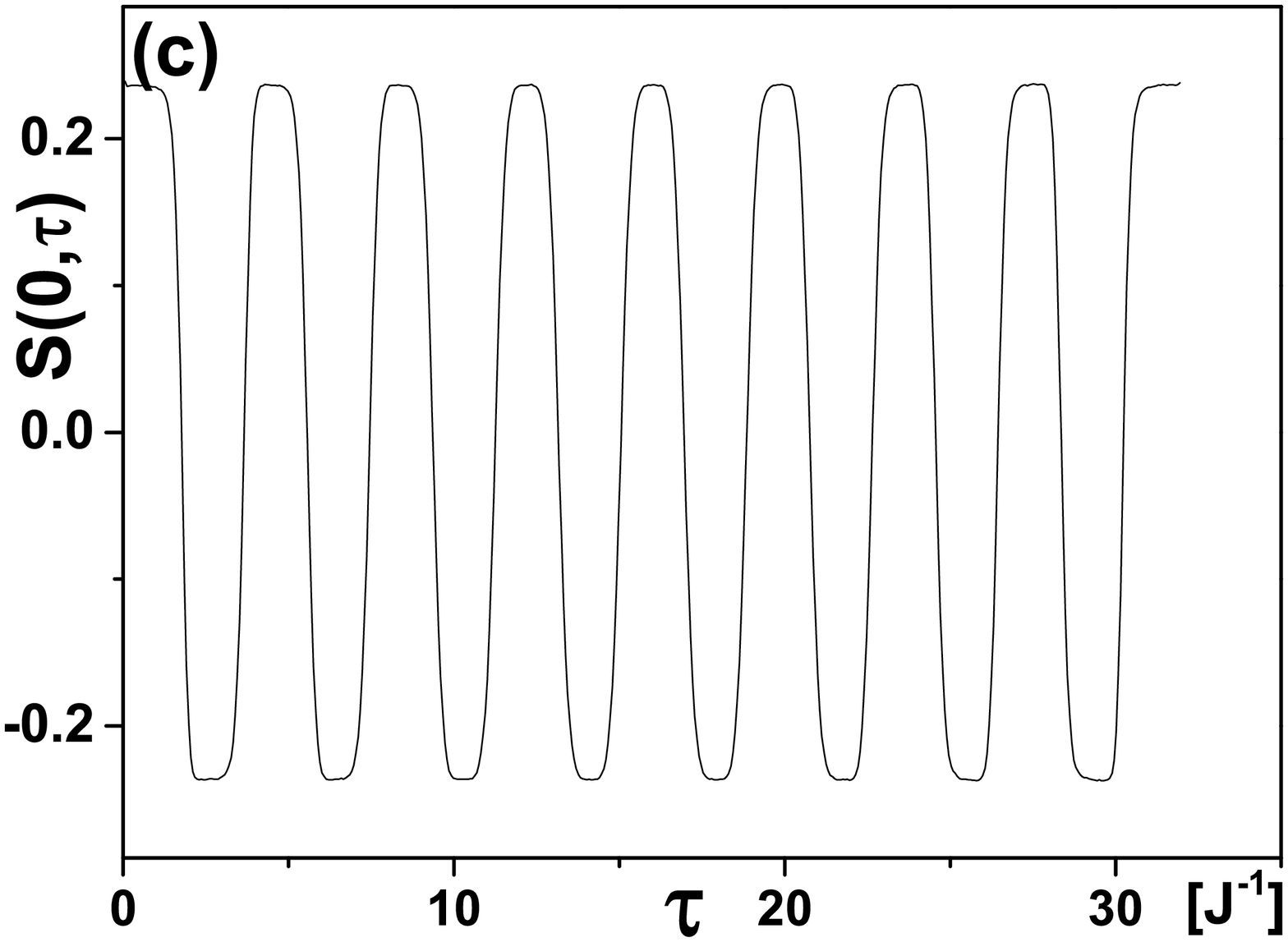}
\caption{(a) The dual transformation between a world line configuration and the
  gas   of interacting point-like particles (i.e., the gas of instantons and
  anti-instantons).  (b) The interaction kernel
  function in the world line configurations (upper panel) and
 instanton  configuration (lower panel).  (c) Onsite density correlation  as a
  function of $\tau$ in the presence of the retarded interaction with the
  kernel~(\ref{eq:ret2}) and parameters $\alpha=6J$, $\sigma=0.5/J$,
  $\tau_0=2/J$, and $\beta J=L=32$.}  \label{fig:fig2}
\end{figure*}

We further assume that  the system particles (i.e., hard-core
  bosons) couple to the bath quasiparticles locally via its density operator
  $n_i$:
$H_{sb}=\sum_{i,\sigma} n_i (\alpha_{i,\sigma}+\alpha_{i,\sigma}^\dag)$. By
integrating out the bath degrees of freedom, one can obtain the bath-induced
retarded interaction between the system particles:
$S_{ret}=\sum_m D(\omega_m) n(\omega_m)n(-\omega_m)$, where the kernel function
takes the form of
$D(\omega_m)\propto \frac {1}{\det[-i\omega_m\hat{1}-\hat{H}]}=\frac
1{(\omega_m-\Gamma)^2+\omega_d^2}$  with  $\Gamma$ and $\omega_d$ defined the same as in Eq.(\ref{M0}).

By performing the Fourier transformation again and take the zero temperature
limit, we obtain $S_{ret}=\int d\tau d\tau' D(\tau-\tau') n(\tau) n(\tau')$, with
 \begin{eqnarray}
 D(\tau-\tau')&\propto& \Re \big[\int d\omega \frac{e^{i\omega(\tau-\tau')}}{(\omega-\Gamma)^2+\omega_d^2}\big]\\
 &\propto& e^{-\omega_d|\tau-\tau'|} \cos\Gamma(\tau-\tau')
 \end{eqnarray}
which is the kernel function shown in the main text.

\section{An alternative model for iTC}
\label{sec:alternative}
In the main text, we choose an oscillating decaying retarded interaction in
imaginary time. However, this oscillating
decaying feature of the retarded interaction is not a necessary condition for
the iTC phase, which is found to exist in a much broader circumstances. That is,
it can exist as long as the
retarded interaction is non-monotonic in $|\tau-\tau'|$, and has at least one
minimum whose position defines the lattice constant of iTC along
$\tau-$direction, similar to the conventional crystals in space.

In this section, we will numerically verify this point by showing the existence of the iTC phase with a different retarded interaction with the kernel function:
\begin{equation}
D(\tau-\tau')=-\frac
{\alpha}{\sigma^2}\big[1-\frac{(|\tau-\tau'|-\tau_0)^2}{\sigma^2}\big]e^{-\frac{(|\tau-\tau'|-\tau_0)^2}{2\sigma^2}}
\,. \label{eq:ret2}
\end{equation}
The physics with such a complex form of the kernel function can be easily
understood by making analogy with the conventional spatial crystals, where the
interacting objects are point-like. However, in the iTC, the interacting objects
are continuous lines (world-line).  Thus to make such an analogue, one needs first
to do a dual transformation,   given world line configuration can be mapped to a
particle picture, i.e., the gas of instantons and anti-instantons (also referred
in literature as to kinks and anti-kinks)  with the same interacting
energy~\cite{Nagaosa1999}:
\begin{equation}
\int d\tau d\tau' \delta n(\tau)D(\tau-\tau')\delta n(\tau')
=\sum_{ij} m(\tau_i) V(\tau_i-\tau_j) m(\tau_j)  \,,
\end{equation}
where $m(\tau_i)=\pm 1$ indicates a kink or antikink with the position $\tau_i$. The interacting energy $V(\tau-\tau')$  between a pair of kinks can be directly evaluated by performing integral over $\tau$ and $\tau'$, which gives rise to
\begin{equation}
V(\tau_i-\tau_j)=\alpha e^{-\frac{(|\tau_i-\tau_j|-\tau_0)^2}{2\sigma^2}} \,.
\end{equation}
Obviously, for those point-like objects (kinks and anti-kinks),  this kind of interaction favor a crystalline pattern in $\tau$-direction with lattice constant $\tau_0$, which corresponds to the iTC pattern in the world-line configurations.

To incorporate the quantum fluctuations in this picture, we perform QMC
simulations of the model that is similar to what is discussed in
  the main text
(hopping+retarded interaction), the only difference being that the kernel
function replaced by Eq.~(\ref{eq:ret2}). In Fig.~\ref{fig:fig2}(c), we plot the
onsite  temporal   density-density correlations,  which shows a
long-range crystalline order in $\tau$-direction for a strong retarded
interaction ($\alpha=6J$), indicating a spontaneous translational symmetry
breaking in imaginary time. The lattice constant of the iTC is determined by
the characteristic  itime length  (the minimum position) of the retarded interaction
($\tau_0$).  This example indicates that the oscillating feature of the retarded
interaction is not necessary for the iTC phase,  and hence proves
  that the phase   can be observed in much
boarder situations.


%

\end{document}